\documentclass[%
aps,prl,
superscriptaddress,
nolongbibliography,
reprint,%
]{revtex4-1}

\usepackage{graphicx}
\usepackage{dcolumn}
\usepackage{bm}
\usepackage{xcolor}
\usepackage{mathrsfs,amsmath}
\usepackage[
			colorlinks = true,
            linkcolor = blue,
            urlcolor  = blue,
            citecolor = blue,
            anchorcolor = blue
            ]{hyperref}         
\newcommand\rxout{\bgroup\markoverwith{\textcolor{red}{\rule[.5ex]{2pt}{.6pt}}}\ULon}

\begin{document}

\title{Extremely asymmetrical acoustic metasurface mirror at the exceptional point}
\author{Xu Wang}
\thanks{X.W. and X.F. contributed equally to this work.}
\author{Xinsheng Fang}
\thanks{X.W. and X.F. contributed equally to this work.}
\author{Dongxing Mao}
\affiliation{Institute of Acoustics, Tongji University, Shanghai 200092, China}
\author{Yun Jing}
\email{yjing2@ncsu.edu}
\affiliation{Department of Mechanical and Aerospace Engineering, North Carolina State University, Raleigh, North Carolina 27695, USA}
\author{Yong Li}
\email{yongli@tongji.edu.cn}
\affiliation{Institute of Acoustics, Tongji University, Shanghai 200092, China}
\affiliation{College of Architecture and Urban Planning, Tongji University, Shanghai 200092, China}

\date{\today}

\begin{abstract}
Previous research has attempted to minimize the influence of loss in reflection- and transmission-type acoustic metasurfaces. This letter shows that, by treating the acoustic metasurface as a non-Hermitian system and by harnessing loss, unconventional wave behaviors that do not exist in lossless metasurfaces can be uncovered. Specifically, we theoretically and experimentally demonstrate a non-Hermitian acoustic metasurface mirror featuring extremely asymmetrical reflection at the exception point. As an example, the metasurface mirror is designed to have high-efficiency retro-reflection when the wave incidents from one side and near-perfect absorption when the wave incidents from the opposite side. This work marries conventional gradient index metasurfaces with the exceptional point from non-Hermitian systems, and paves the way for identifying new mechanisms and functionalities for wave manipulation.  
\end{abstract}

\maketitle

Molding the flow of acoustic energy using functional materials is a research area that has recently generated a proliferation of work \cite{Liu2000Science,Fang2006NM,Brunet2015NM,Yang2008PRL,Cummer2016NRM,Assouar2018NRM}. As a member of functional acoustic materials, acoustic metasurfaces stand out as a distinct choice for wave manipulation owing to their advanced capabilities on sound control as well as their vanishing size \cite{Assouar2018NRM,Li2013SR,Li2014PRApplied,Xie2014NC,Li2015PRApp}. Conventional transmission- or reflection-type acoustic metasurfaces operate by modulating the real part of their effective refractive indices \cite{Li2013SR,Li2014PRApplied,Ma2014NM,Xie2014NC,Li2015PRApp,Zhu2016NC,Xie2017AM} and are typically treated as lossless systems. However, due to the existence of resonance or narrow regions in the deep-subwavelength units, losses are naturally present \cite{Li2016APL,Moleron2016NJP}, rendering non-zero imaginary part of the refractive index. The intrinsic loss could compromise the performance of acoustic metasurfaces and the conventional wisdom is that their effects should be minimized. The emergence of non-Hermitian physics \cite{Bender1998PRL}, however, offers a brand new prospective on the role of loss. Instead of minimizing the loss in functional materials, recent research suggests that losses can be harnessed to engender highly unusual phenomena \cite{Li2017PRL}. 

The publication of the seminal paper by \citet{Bender1998PRL} immediately spurred an intense interest in quantum mechanics on non-Hermitian Hamiltonian. Their theory describes a new family of systems that, though violate time-reversal ($\cal{T}$) symmetry, retain the combined parity-time ($\cal{PT}$) symmetry. Such a system possesses entirely real-valued energy spectra below the ${\cal P}{\cal T}$ symmetry breaking threshold -- the exception point (EP), where the associated eigenvalues and the corresponding eigenvectors coalesce \cite{Bender1999JMP,Miri2019Sci}. Introducing $\cal{PT}$ symmetry into the classical optic and mechanical wave systems has paved the way for identifying new mechanisms to control light and sound \cite{Ruter2010NP,Shi2016NC,Auregan2017PRL,Fleury2015NC,Zhu2014PRX}. Given that gain is more challenging to achieve than loss in practical systems, it has been proposed to relax the restriction on exact gain/loss modulation, giving rise to the so-called passive non-Hermitian $\cal{PT}$-symmetric systems -- systems with only loss \cite{Feng2013NM,Shen2018PRM,Liu2018PRL}. In fact, besides $\cal{PT}$ symmetric systems, EPs can be also observed in other non-Hermitian systems \cite{Zhu2018PRL,Feng2014OE}. In these passive systems, intriguing features such as unidirectional near-zero reflection and unidirectional focusing have been observed at the EP \cite{Feng2013NM,Liu2018PRL,Shen2018PRM,Zhu2018PRL}.

\begin{figure}
\includegraphics[width=8.6cm]{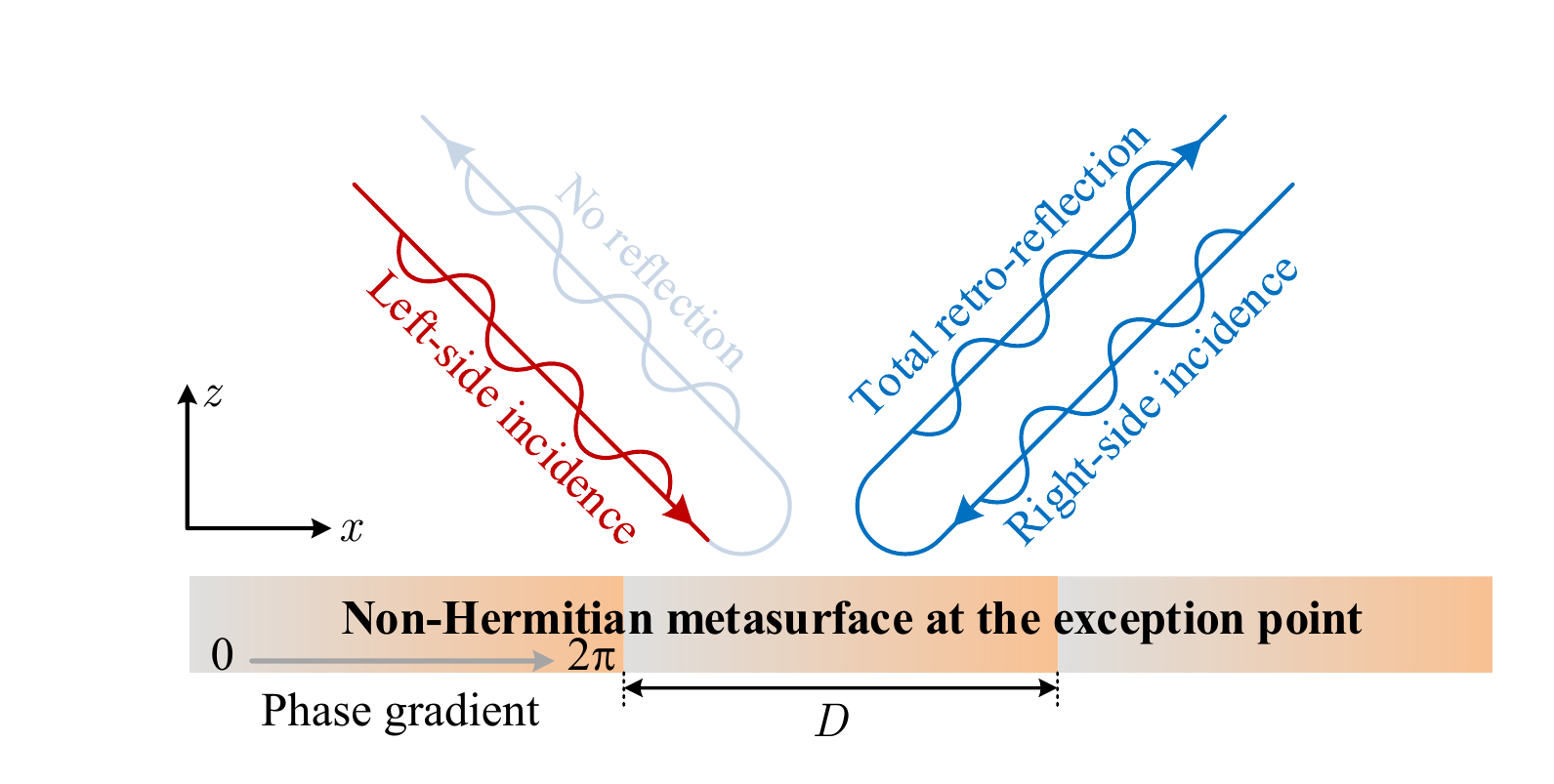}
\caption{\label{fig:1}Schematic of the non-Hermitian metasurface showing extremely asymmetrical reflections at the EP.}
\end{figure}

Previous works, however, are largely based on the archetype \cite{Ruter2010NP,Shi2016NC,Auregan2017PRL,Feng2013NM,Shen2018PRM,Zhu2018PRL} or modification (i.e. by curving or extending) \cite{Zhu2014PRX,Liu2018PRL} of the 1D waveguide model, which is a 1D scattering system where the wave propagates along modulated potentials. Studies on higher dimensions are scarce and mostly theoretical \cite{Ge2014PRX,Turduev2015PRA} and this is also true for studies on EPs in metasurfaces \cite{Kang2016PRA,Sakhdari2017NJP}. In this letter, by marrying acoustic metasurfaces and the concept of EP, we theoretically and experimentally demonstrate a non-Hermitian metasurface mirror in 2D space. In this system, the elegantly engineered loss provides an additional degree of freedom, from which extremely asymmetrical yet arbitrarily-tailored reflection can be observed at the EP.
\begin{figure*}[tb!]
\includegraphics[width=14cm]{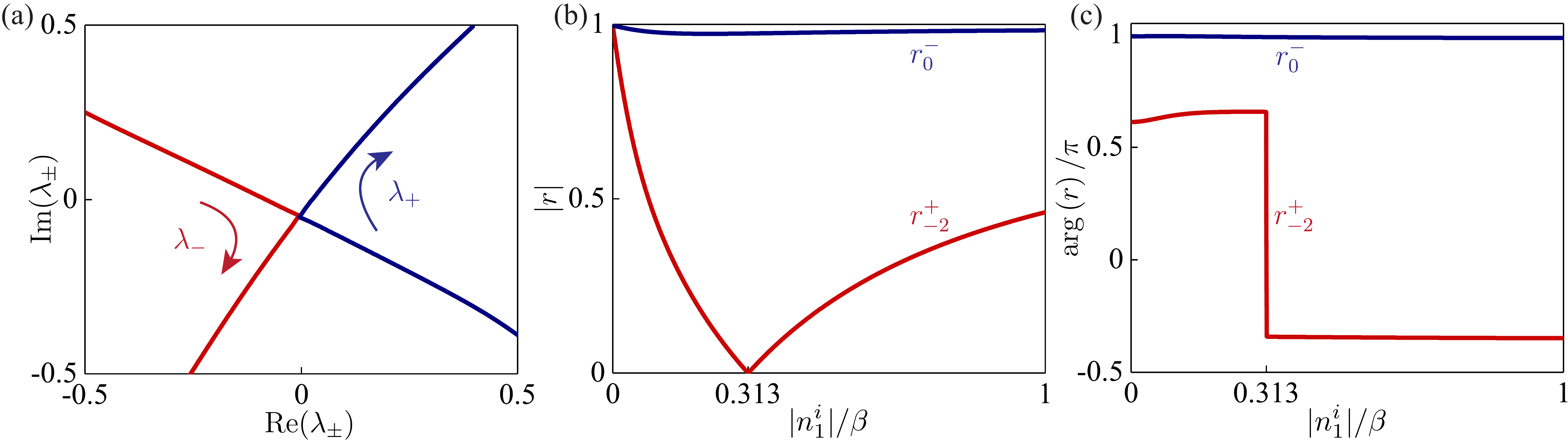}
\caption{\label{fig:2}(a) The trajectories of eigenvalues ${\lambda _ \pm }$ of the scattering matrix in the complex plane by varying the loss of unit 1 (${n_{1}^{i}}$). (b) Amplitude and (c) phase of extraordinary reflection coefficients $r_0^ - /r_{ - 2}^ + $ by varying ${n_{1}^{i}}$. }
\end{figure*}
Additionally, a crux in designing passive, unidirectional reflectionless systems \cite{Feng2013NM,Liu2018PRL,Shen2018PRM,Feng2014OE} is that, while it is relatively easy to have the reflection from one side diminished, it is extremely challenging to, at the same time, have high reflectivity from the opposite side. For instance, a recent work has demonstrated asymmetrical reflection for electromagnetic waves \cite{Wang2018PRL}. To realize extreme asymmetry (reflectivity from one side being over $90\% $ while that from the opposite being close to zero) though, more than 100 sub-units need to be included in one period, and all sub-units must be elaborately modulated in index and loss \cite{Wang2018PRL}. 
In contrast, by leveraging EP, our metasurface works effectively with sparse sub-units and remarkably, only one sub-unit is required to be lossy. As will be shown in the following, such a simple design principle could in theory achieve over $97\% $ reflectivity from one side and zero reflectivity from the other side. 

A gradient index acoustic metasurface, as depicted in Fig.~\ref{fig:1}(a), is studied. This metasurface consists of N sub-units in one period, each having an effective refractive index ${n_i}$ (i=1, 2, ..., N). The reflected waves are constituted by different diffraction modes, whose angles satisfy \cite{Yu2011Science,Xie2014NC,Li2017PRL}
\begin{equation}
{k_0}\left( {\sin {\theta _r} - \sin {\theta _i}} \right) = \xi  + mG, \label{eq:1}
\end{equation} 
where ${k_0}$ is the wave number, ${\theta _i}/{\theta _r}$ represents the angle of incidence/reflection, $\xi $ is the phase gradient induced by the metasurface ($\xi  = d\varphi /dx$, $\varphi $ is the phase shift, and $x$ is the coordinate along the surface), $G$ is the amplitude of the reciprocal lattice vector ($G = 2\pi /D$, $D$ is the period of the metasurface), and $m$ is the diffraction order.
\begin{figure*}
\includegraphics[width=14cm]{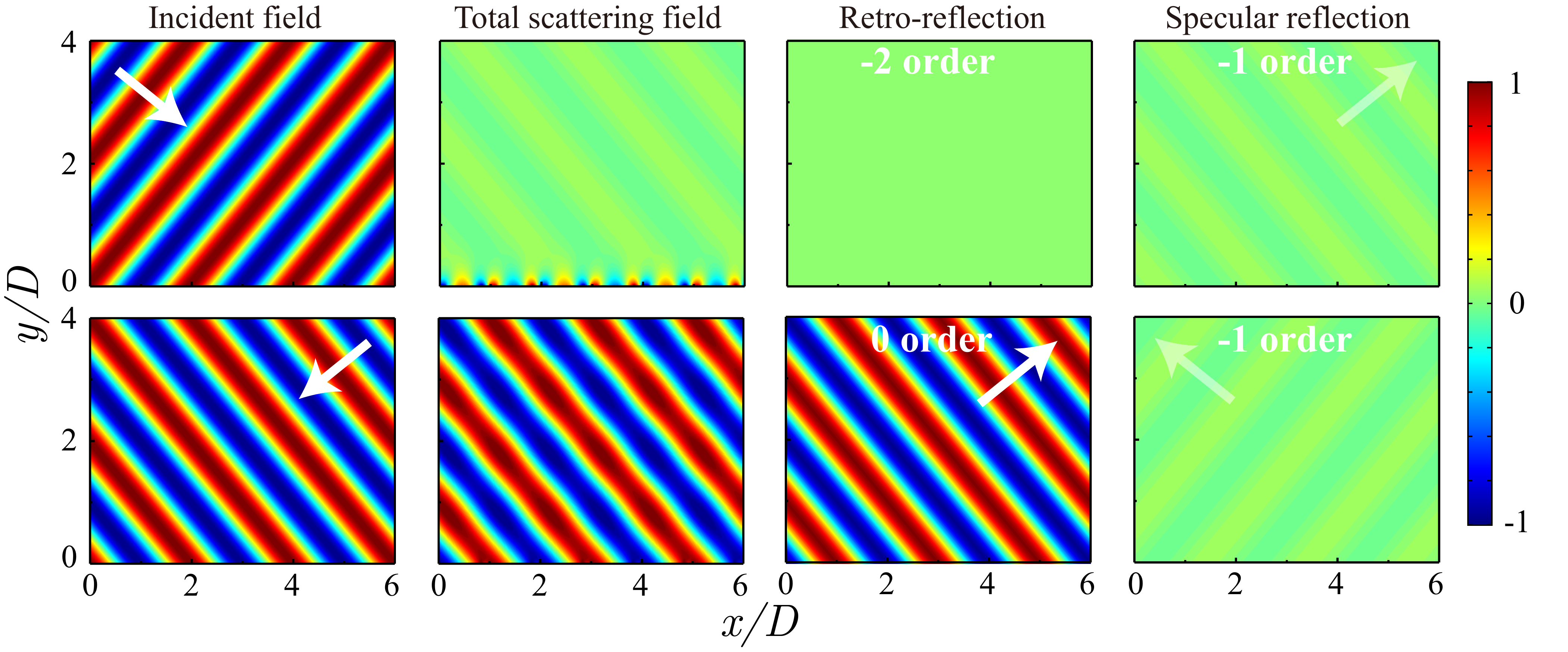}
\caption{\label{fig:3}The upper figures shows the theoretical predicted pressure field with left-side incidence (${\theta _i} = {45^ \circ }$), from left to right: incident field, total scattered field (contributed by all diffraction modes), retro-reflection (contributed by diffraction order of -2), and specular reflection (contributed by diffraction order of -1); the lower figures shows the theoretical predicted pressure field with right-side incidence (${\theta _i} =  - {45^ \circ }$), from left to right: incident field, total scattered field (contributed by all diffraction modes), retro-reflection (contributed by diffraction order of 0), and specular reflection (contributed by diffraction order of -1).}
\end{figure*}

In order to introduce the EP, the metasurface is designed as a two-port system so that only two diffraction modes could propagate while others are suppressed. When the acoustic wave at the designated angle of incidence impinges upon the metasurface, it splits into two parts: an extraordinary reflection (we choose retro-reflection in this letter as an example) and an ordinary one (specular reflection). Without loss of generality, the angle of incidence is set as ${\theta _i} =  \pm {45^ \circ }$ for the left-/right-side incidence (see Section A of Supplemental Material \cite{SupplNote} for the design at other angles). In order to satisfy the requirement of our design, the reciprocal lattice vector is set to $G = \sqrt 2 {k_0}$ (equivalent to $D = {\lambda _0}/\sqrt 2 $). For the left-side incidence (${\theta _i} = {45^ \circ }$), as predicted by Eq.~(\ref{eq:1}), only the diffraction orders of $m =  - 1$ and $m =  - 2$ give rise to propagating waves, corresponding to the specular reflection (${\theta _r} = {45^ \circ }$) and the extraordinary reflection (retro-reflection with ${\theta _r} =  - {45^ \circ }$), respectively. For the right-side incidence (${\theta _i} =  - {45^ \circ }$), only the diffraction orders of $m =  - 1$ and $m =  0$ produce propagating waves, representing the specular (${\theta _r} =  - {45^ \circ }$) and extraordinary (${\theta _r} = {45^ \circ }$) reflections, respectively.    
 
Such a two-port system can be described by its scattering matrix $S$ \cite{Zhu2014PRX}, which connects the obliquely forward-/backward-going waves with the opposite-side incidence (illustrated by Fig.~\ref{fig:1}), as
\begin{equation}
\left[ \begin{array}{l}
p_f^R\\
p_b^L
\end{array} \right] = S\left[ \begin{array}{l}
p_f^L\\
p_b^R
\end{array} \right],  \label{eq:2}
 \end{equation} 
where $p$ is  the complex pressure amplitude, the subscript $f/b$ indicates forward-/backward-going waves and the superscript $L/R$ indicates the left-/right-side of the surface normal of the incident location. For example, $p_f^L$ is the left-side incidence; $p_f^R$ represents both the specular reflection under left-side incidence and retro-reflection under right-side incidence (these two paths overlap in our design). Consequently, the scattering matrix $S$ reads
\begin{equation}
 \label{eq:3} 
S = \left[ \begin{array}{l}
\begin{array}{*{20}{c}}
{r_{ - 1}^ + }&{r_0^ - }
\end{array}\\
\begin{array}{*{20}{c}}
{r_{ - 2}^ + }&{r_{ - 1}^ - }
\end{array}
\end{array} \right].
\end{equation}  
Here $r$ is the reflection coefficient of the corresponding diffraction mode. The subscript indicates the mode order $m$, and the superscript $ + / - $ indicates the left-/right-side incidence. In this letter, by using the coupled-mode theory (see Section B of Supplemental Material \cite{SupplNote}), we can obtain the complete reflectivity spectra of the metasurface mirror. 

Owing to reciprocity, specular reflections from both sides are identical, i.e.,$r_{ - 1}^ +  = r_{ - 1}^ - $. The left-side extraordinary reflection $r_{ - 2}^ + $, however, can be different from that of the opposite side, $r_0^ - $. Here, the designed metasurface consists of 6 sub-units per period. Loss is introduced to sub-unit 1 (can be any other sub-unit), while other units are lossless. This means that only sub-unit 1 has a complex refractive index (${n_1} = {n_{1}^{r}} + i{n_{1}^{i}}$ and ${n_{1}^{i}} < 0$). The asymmetry of the metasurface mirror, therefore, can be explored by varying the loss of sub-unit 1, i.e. ${n_{1}^{i}}$. It should be pointed out that, in order to maximize the retro-reflection efficiency for the right-side incidence, the refractive indices are fine-tuned around the exact values determined by the index gradient (see Section C of Supplemental Material \cite{SupplNote} for the two-step method for designing the metasurface mirror). This is a common practice in optimizing the performance of metasurfaces \cite{Rubio2017PRB}.

The eigenvalues ${\lambda _ \pm } = r_{ - 1}^ +  \pm \sqrt {r_{ - 2}^ + r_0^ - } $ of the scattering matrix are calculated using the coupled mode theory. Their trajectories \cite{Gear2017NJP} in the complex plane are shown in Fig.~\ref{fig:2}(a), featuring anticrossing real parts and crossing imaginary parts -- a hallmark of EPs \cite{Huang2015OE}. At the EP where ${n_{1}^{i}}$ reaches $-0.313 \beta $ (${\beta  = {{\lambda _0}}/{{2Nd}}}$ and $d$ is the metasurface thickness), the two eigenvalues  ${\lambda _ \pm }$ coalesce, together with the corresponding eigenvectors ${v_ \pm } = {\left[ {\begin{array}{*{20}{c}}
1&{ \pm \sqrt {r_{ - 2}^ + /r_0^ - } }
\end{array}} \right]^T}$. This is a direct manifestation of the EP from the non-Hermiticity of the system \cite{Huang2015OE}.
\begin{figure*}
\includegraphics[width=13.5cm]{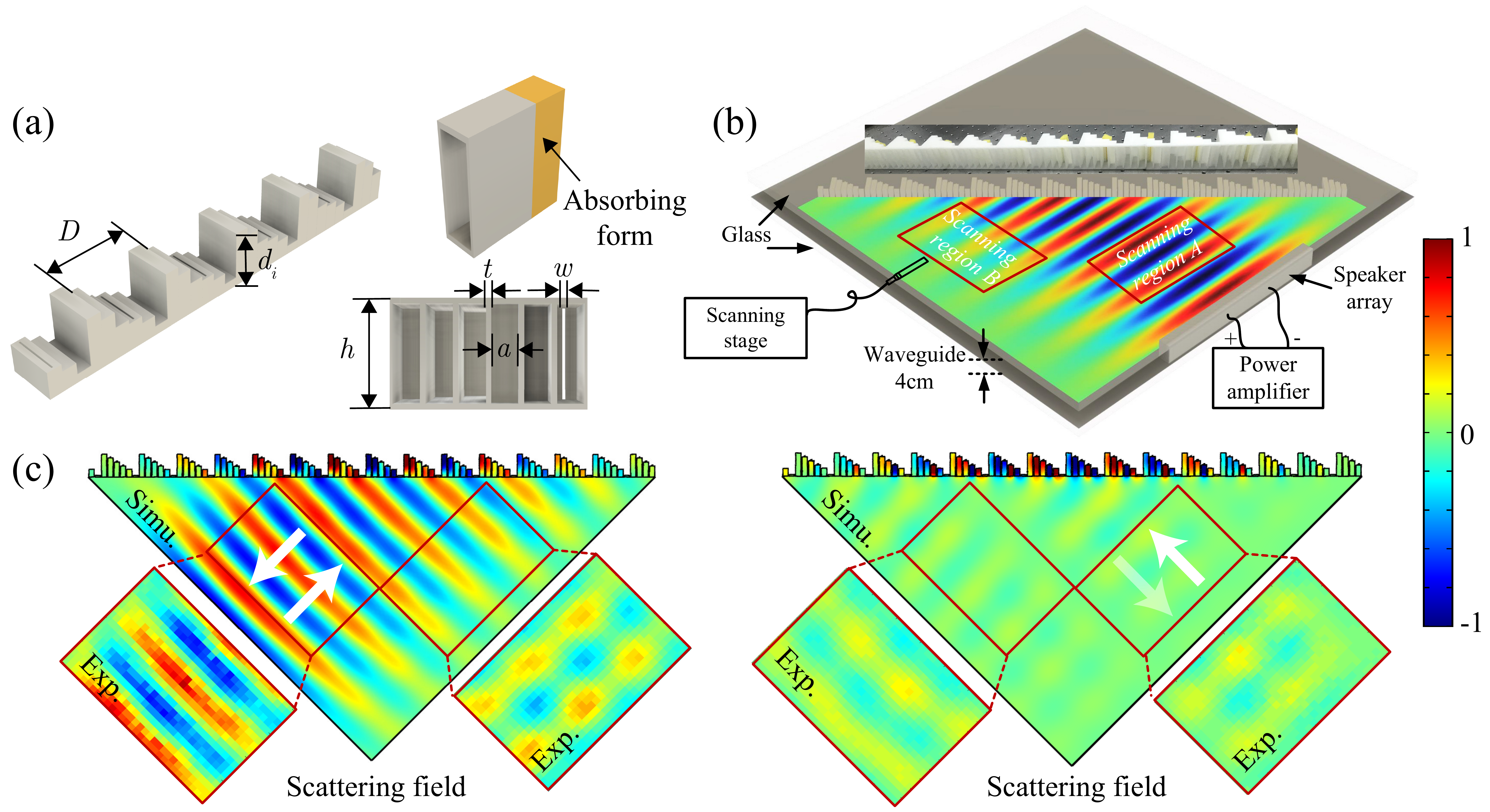}
\caption{\label{fig:4}Illustration of the metasurface mirror constructed by groove and leaky-groove sub-units (left). The sample is flipped-over to show the bottom of the structure. The leaky unit is attached by an absorbing foam (top right), and bottom right shows the top view of one period of the sample. (b) Experimental setup. (c) Simulated and experimental results (reflected field) of the metasurface mirror at the EP.}
\end{figure*}
The theoretically calculated amplitudes and phases of these two extraordinary reflection coefficients ($r_{ - 2}^ + $ and $r_0^ - $) are shown in Figs.~\ref{fig:2}(b) and \ref{fig:2}(c), respectively. The extraordinary reflections from opposite sides respond very differently to the varying loss. When the loss of unit 1 is zero, total extraordinary reflections arise from both sides. As the loss is initially increased, left-side retro-reflection ($r_{ - 2}^ + $) has a declined amplitude. A peculiar case where the left-side retro-reflection vanishes can be observed at the EP when ${n_{1}^{i}}$ reaches $-0.313 \beta$. Meanwhile, the right-side retro-reflection is very strong ($\left| {r_0^ - } \right| = 0.974$). As shown by Fig.~\ref{fig:2}(c), the phase of the left-side retro-reflection ($r_{ - 2}^ + $) experiences an abrupt change of $\pi $ at the EP. 
Interestingly, $\left| {r_{ - 2}^ + } \right|$ is found to increase with the loss beyond EP. This unconventional phenomenon, dubbed loss enhanced reflection, is very similar to the loss enhanced transmission reported in Ref.~\cite{Guo2009PRL}. On the other hand, the right-side retro-reflection is very stable against the varying loss, maintaining a very high efficiency ($\left| {r_0^ - } \right| > 0.97$) and a virtually constant phase shift [$\arg \left( {r_0^ - } \right) = \pi $]. 

We theoretically calculate the acoustic pressure field of the metasurface mirror at the EP. The top/bottom figures in Fig.~\ref{fig:3} illustrate the left-/right-side incidence case. From left to right, they show the incident field, the total scattered field contributed by all diffraction modes, the component of retro-reflection contributed by $r_{ - 2}^ + /r_0^ - $, and the component of specular reflection contributed by $r_{ - 1}^ + /r_{ - 1}^ - $, respectively. Figures in the second column manifest the asymmetrical response of the metasurface to opposite incidences. Compared to the left-side incidence that generates strong evanescent waves, the waves reflect straight back under the right-side incidence. Due to reciprocity, specular reflections show a symmetrical pattern as indicated by the last column of Fig.~\ref{fig:3}. Here $\left| {r_{ - 1}^ + } \right| = \left| {r_{ - 1}^ - } \right| = 0.048$. This part of energy flux with an intrinsically symmetric behavior is strongly suppressed so that the overall asymmetrical behavior can be enhanced. The figures in the third column further demonstrate the EP of the system, where no retro-reflection is found under left-side incidence ($\left| {r_{ - 2}^ + } \right| = 0$) and a strong retro-reflection is present under right-side incidence ($\left| {r_0^ - } \right| = 0.974$).


To validate the theory, corresponding experiments and numerical simulations are carried out. Inspired by the work of \cite{Zhu2018NC}, the metasurface mirror is constructed by periodically arranged grooves, one of which is leaky (in order to introduce loss) and is covered by a layer of absorptive material at the end [Fig.~\ref{fig:4}(a)].  By tuning the depth of the groove ($d$) and the width ($w$) of  the slit, this structure, albeit very elementary, provides fine yet arbitrary manipulation on the real and imaginary parts of the effective refractive index (see Section D of Supplemental Material \cite{SupplNote}). The operating frequency is 3430 Hz and the period of the metasurface is $D=7.07$ cm (${\lambda _0}/\sqrt 2 $). In each period, 6 grooves are used with the following depths ($d_i$): 2.19, 2.56, 3.85, 4.60, 0.59, and 1.57 cm. All units have the same groove width ($a=9.79$ mm), same wall thickness ($t=1$ mm), and only unit 1 has a slit at its end with a slit width of $w=1.34$ mm.

The performance of the metasurface is first investigated by COMSOL Multiphysics. For the measurement, a two-dimensional waveguide with a height of $h=4$ cm is built where the acoustic field can be scanned. To scan the field, a microphone moves at a step size of 1 cm in the two regions marked by red boxes in Fig.~\ref{fig:4}(b) (which identify the retro-reflection and specular reflection regions). The length of the sample is 1.06 m (15 periods). The scattered field is obtained by subtracting the incident field (scanned without the metasurface) from the total field (scanned with the metasurface). The normalized simulated and measured results are presented in Fig.~\ref{fig:4}(c). Excellent agreement can be found between the simulated and measured results, both suggesting a strongly asymmetrical wave behavior. For the right-side incidence (figure on the left), retro-reflection is clearly observable, and the estimated pressure reflection coefficient reaches around 0.87 (simulation) and 0.83 (measurement), respectively. Both reflectivities are lower than predicted possible due to the mutual coupling effect between sub-units that is not considered in the theoretical model. Under the opposite incidence, such a reflection is strongly suppressed at the EP due to the judiciously designed loss. It should be pointed out that, although the EP is achieved at 45$^ \circ $ for the current design, strongly asymmetrical behavior is in fact observable under a wide range of angles of incidence (see Section E of Supplemental Material \cite{SupplNote}). 




To conclude, we have theoretically studied and experimentally validated a non-Hermitian acoustic metasurface mirror. We show that the 2D extreme asymmetry in terms of the extraordinary reflection can be interpreted by the EP of the non-Hermitian system. This peculiar behavior not found in conventional metasurfaces hinges on elegantly-engineered losses, which provide an additional degree of freedom for wave manipulation. Our findings expand the exploration of acoustic metasurfaces into the complex domain, and may open a new route for wave manipulation with new functionalities.

\begin{acknowledgements}
	This work was supported by the National Natural Science Foundation of China (Grant Nos. 11704284 and 11774265), the Young Elite Scientists Sponsorship by CAST (Grant No. 2018QNRC001), the Shanghai Science and Technology Committee (Grant No. 18JC1410900), the Shanghai Pujiang Program (Grant No. 17PJ1409000), and the Stable Supporting Fund of Acoustic Science and Technology Laboratory.
\end{acknowledgements}

%

\end{document}